\documentclass[aps,floatfix,prb,twocolumn,showpacs]{revtex4-1} 
\usepackage{graphicx}
\usepackage{amsmath}
\usepackage{MnSymbol}
\begin{document}
\title{Exact solution for Bloch oscillations of a simple charge-density-wave insulator} 
\author{W. Shen$^1$}
\author{ T. P. Devereaux$^{2,3}$}
\author{ J. K. Freericks$^1$ }
\address{$^1$Department of Physics Geogetown University,
Washington, DC 20057,USA\\
$^2$Stanford Institute for Materials and Energy Sciences, SLAC National Accelerator Laboratory, Stanford University, Stanford, CA 94305, USA\\
$^3$Geballe Laboratory for Advanced Materials, Stanford University, Stanford, CA 94305, USA
} 
\date{\today} 

\begin{abstract}
Charge-density-wave systems have a static modulation of the electronic charge at low temperatures when they enter an ordered state. While they have been studied for decades in equilibrium, it is only recently that they have been examined in nonequilibrium with time-resolved studies. Here, we present the exact solution for the nonequilibrium response of electrons (in the simplest model for a charge density wave) when the system is placed under a strong DC electric field. This allows us to examine the formation of driven Bloch oscillations and how the presence of a current modifies the nonequilibrium density of states.
\end{abstract}
\maketitle

\section{Introduction}
Charge-density-wave (CDW) behavior occurs in a wide variety of different materials. Peierls\cite{peierls} originally showed the instability of a one-dimensional metal to a distortion with a unit cell twice as large as in the uniform phase, that opens an insulating gap at the Fermi level. Since then, CDW behavior has been seen in many different materials and in higher dimensions as well. Still, the precise origin of CDW behavior in real materials is controversial\cite{mazin,kivelson,tpd_cdw}---is it arising from electron-phonon interactions and a softening phonon or from a purely electronic instability in the static charge susceptibility or via an instability driven by the electron-phonon coupling? We do not investigate that question here, but instead focus on the behavior in the nonequilibrium state.

Recently, nonequilibrium  photoelectron pump-probe experiments have been carried out for a number of different CDW systems, with both valence electron photoemission\cite{perfetti1,perfetti2,rossnagel,felix,cavalleri} and core-hole photoemission\cite{rossnagel2}.  In these experiments,  the charge density wave material is pumped with an infrared laser pulse and displays nonequilibrium melting of the CDW state, which is illustrated by a filling in of the gap in the photoemission spectrum, while the system still retains its modulation of the electronic charge in the ordered CDW phase. This phenomena has already been examined with an exactly solvable model\cite{tr-pes-us}.

The simplest model for a CDW insulator is to start with a system that can be divided into two sublattices, called $A$ and $B$ and having hopping only between the two sublattices.  Then we pick an on-site energy to be equal to $U$ on the $A$ sublattice and 0 on the $B$ sublattice. The equilibrium Hamiltonian becomes
\begin{equation}
\mathcal{H}=-\displaystyle\sum_{ ij }t_{ij}\emph{c}_{i}^{\dagger}\emph{c}_{j}^{}+\displaystyle\sum_{i\in{A}}(U-\mu)\emph{c}_{i}^{\dagger}\emph{c}_{i}^{}+\displaystyle\sum_{i\in{B}}(-\mu)\emph{c}_{i}^{\dagger}\emph{c}_{i}^{}.
\label{eq: ham_eq}
\end{equation}
Here $\emph{c}_{i}^{\dagger}$ and $\emph{c}_{i}^{}$ are the creation and annihilation operators for a spinless fermion at site $i$. The operators satisfy the canonical anticommutation relations
\begin{equation}
\{c_i^{}, c_j^{\dagger}\}_+=\delta_{ij},
\end{equation}
and
\begin{equation}
\{c_i^{}, c_j^{}\}_+=\{c_i^{\dagger}, c_j^{\dagger}\}_+=0
\end{equation}
where the $+$ subscript denotes the anticommutator of the two operators.
In Eq.~(\ref{eq: ham_eq}), $\mu$ is the chemical potential and $U$ is the aforementioned site energy. The electrons are allowed to hop between nearest neighbors with a hopping matrix $-t_{ij}$, which is a real and symmetric matrix that equals $-t$ for $i$ and $j$ nearest neighbors and vanishes otherwise.
When this Hamiltonian is diagonalized (see below), it forms two bands, so filling the electrons halfway (one electron per two lattice sites), yields an insulating phase.  Because the site energy is fixed and never varies, the CDW order is always frozen in, and remains for all finite temperatures. While this might seem like an extreme limit, it should describe the behavior of experimentally studied CDW systems for short times, before the phonons are able to relax the system and reduce or eliminate the effective site energy (which arose from the phonon distortion in the ordered phase).  Nevertheless, because this model can be solved exactly, it provides an interesting limit for other (more accurate) model calculations and displays interesting nonequilibrium behavior that has a number of nontrivial results.

The organization of this paper is as follows. In Section II, we develop the formalism for the exact solution of the nonequilibrium problem. In Section III, we present our solutions for the Bloch oscillations and nonequilibrium density of states, and we conclude in Section IV.

\section{Formalism} 

In this section, we will first describe how to solve for the equilibrium bandstructure of the CDW and then we will show how to employ the Peierls substitution to describe the nonequilibrium solution.

\subsection{Equilibrium formalism}

The lattice translational symmetry is broken when $U$ is nonzero, resulting in a doubling of the size of the unit cell in real space, and a halving of the Brillouin zone in reciprocal space. Hence, the conversion from real space to momentum space is more complicated than in a system with one atom per unit cell. The
momentum points ${\bf k}$ and ${\bf k}+{\bf Q}$ are coupled, where ${\bf Q}=(\pi,\pi, \dots )$ due to the presence of the CDW order. The transformation from reciprocal space to real space becomes
\begin{equation}
\emph{c}_{i}^{\dagger}=\displaystyle\sum_{k}(e^{-i\textbf{k}\cdot{\textbf{R}_{i}}}\emph{c}_{k}^{\dagger}+e^{-i(\textbf{k}+\textbf{Q})\cdot{\textbf{R}_{i}}}\emph{c}_{k+Q}^{\dagger}),
\end{equation}
where the sum is over the reduced Brillouin zone, and ${\bf R}_i$ is the position vector for the $i$th lattice site.
Since $e^{-i\textbf{Q}\cdot\textbf{R}}$ is equal to one for lattice sites on the $A$ sublattice and minus one on the $B$ sublattice, we have explicit expressions
\begin{equation}
\emph{c}_{i\in{A}}^{\dagger}=\displaystyle\sum_{k}e^{-i\textbf{k}\cdot{\textbf{R}_{i}}}(\emph{c}_{k}^{\dagger}+\emph{c}_{k+Q}^{\dagger}),
\end{equation}
\begin{equation}
\emph{c}_{j\in{B}}^{\dagger}=\displaystyle\sum_{k}e^{-i\textbf{k}\cdot{\textbf{R}_{j}}}(\emph{c}_{k}^{\dagger}-\emph{c}_{k+Q}^{\dagger}).
\end{equation}
The corresponding annihilation operator identities are found by taking the respective hermitian conjugates.

If we write the electronic band structure at $U=0$ as $\varepsilon_k$, then we have
\begin{eqnarray}
\varepsilon_k&=&-\displaystyle\sum_{ ij }t_{ij}\exp [-i\textbf{k}\cdot (\textbf{R}_{i}-\textbf{R}_{j})]
\label{eq: bandstructure}\\
&=&-2t\sum_{l=1}^d \cos(k_la)=-\lim_{d\rightarrow\infty}\frac{t^*}{\sqrt{d}}\displaystyle\sum_{l=1}^{d}\cos (k_l a),\nonumber
\end{eqnarray}
where we assumed the hopping was only between nearest neighbors on a $d$-dimensional hypercubic lattice and satisfied $t=t^*/(2\sqrt{d})$ in the limit of large dimensions in the second line (we will present results in the infinite-dimensional limit here, but the formalism gives the exact solution in any dimension).
Restricting $k$ to the reduced Brillouin zone, is equivalent to having $\varepsilon_k\leq 0$. 
In Eq.~(\ref{eq: bandstructure}), $l$ is the index for spatial component along an axial direction, $a$ is the lattice constant, and $d$ is the number of spatial dimensions.

The Hamiltonian in Eq.~(\ref{eq: ham_eq}) can now be written in a $2\times2$ $\overgroup{\undergroup{c_k, c_{k+Q}}}$ basis via
\begin{equation}
\mathcal{H}= \displaystyle\sum_{k}\undergroup{\overgroup{\begin{array}{cc}\emph{c}_{k}^{\dagger}&\emph{c}_{k+Q}^{\dagger}\end{array}}}
\left(\begin{array}{cc}{\frac{U}{2}-\mu+\varepsilon_k} & {\frac{U}{2}} \\ {\frac{U}{2}} & {\frac{U}{2}-\mu-\varepsilon_k}\end{array}\right)
\left(\begin{array}{c}\emph{c}_{k}\\\emph{c}_{k+Q}\end{array}\right).
\end{equation}
This is diagonalized via the following eigenfunction basis,
\begin{equation}
c_{k+}=\alpha_k\emph{c}_{k}+\beta_k\emph{c}_{k+Q},
\end{equation}
\begin{equation}
c_{k-}=\beta_k\emph{c}_{k}-\alpha_k\emph{c}_{k+Q},
\end{equation}
with $\alpha_k$ and $\beta_k$ satisfying
\begin{equation}
\alpha_k=\frac{\frac{U}{2}}{\sqrt{2\left (\varepsilon_k^2+\frac{U^2}{4}-\varepsilon_k\sqrt{\varepsilon_k^2+\frac{U^2}{4}}\right )}}
\label{eq: alpha_eq}
\end{equation}
and
\begin{equation}
\beta_k=\frac{-\varepsilon_k+\sqrt{\varepsilon_k^2+\frac{U^2}{4}}}{\sqrt{2\left (\varepsilon_k^2+\frac{U^2}{4}-\varepsilon_k\sqrt{\varepsilon_k^2+\frac{U^2}{4}}\right )}}.
\label{eq: beta_eq}
\end{equation}
The operators $c^\dagger_{k+}$ and $c^\dagger_{k-}$ create electrons in the upper and lower bands, respectively.
The Hamiltonian matrix is then diagonalized as follows:
\begin{equation}
\mathcal{H}=\displaystyle\sum_k\left ( \varepsilon_{k+}c_{k+}^\dagger c_{k+}^{}+
\varepsilon_{k-} c_{k-}^\dagger c_{k-}^{}\right).
\end{equation}
Here, $\varepsilon_{k+}$ and $\varepsilon_{k-} $ are given by
\begin{equation}
\varepsilon_{k\pm}=\frac{U}{2}-\mu \pm \sqrt{\varepsilon_k^2+\frac{U^2}{4}}.
\end{equation}

We will also want to work with Green's functions.
The local retarded Green's function is defined by the following formula in equilibrium
\begin{equation}
G^R_{A,B} (t)=-i \theta (t-t')\displaystyle\sum_k\langle \{c_k(t)\pm c_{k+Q}(t) , c_k^\dagger (0)\pm c_{k+Q}^\dagger (0) \}_+\rangle 
\end{equation}
where $\hat{O}(t)=e^{iHt}Oe^{-iHt}$ is the operator representation in the Heisenberg picture and the 
angle brackets are a shorthand for the trace over all states weighted by the density matrix, which
is equal to $\exp[-\beta\mathcal{H}]/(\rm{Tr} \exp[-\beta\mathcal{H}])$, with $\beta$ the inverse temperature (in this work we will start the system from zero temperature or $\beta\rightarrow\infty$).
The plus sign represents  the $A$ sublattice and the minus sign represents the $B$ sublattice. 

In order to explicitly determine the Green's function, it is more convenient to work in the diagonalized basis, where the local retarded Green's function on the $A$ sublattice becomes,
\begin{eqnarray}
G^R_{A} (t)&=&-i \theta (t-t')\sum_k \langle \{(\alpha_k +\beta_k)c_{k+}(t)+(\beta_k -\alpha_k)c_{k-}(t),
\nonumber\\
&~&(\alpha_k +\beta_k)c^{\dagger}_{k+}(0)+(\beta_k -\alpha_k)c^{\dagger}_{k-}(0)\}_+\rangle.
\end{eqnarray}
The time evolution of the $c_{k+}(t)$ and $c_{k-}(t)$ operators is found by solving their equations of motion to yield
\begin{equation}
c^{\dagger}_{k+}(t)=\exp(i\varepsilon_{k+}t)c^{\dagger}_{k+}(0)
\end{equation}
\begin{equation}
c^{\dagger}_{k-}(t)=\exp(i\varepsilon_{k-}t)c^{\dagger}_{k-}(0).
\end{equation}
The local equilibrium density of states (DOS) $A_i(\omega)=-{\rm Im}G^R_i(\omega)/\pi$ then becomes
\begin{equation}
A_{A,B}(\omega)= {\rm Re} \left[ \sqrt{\frac{\omega\pm \frac{U}{2}}{\omega\mp \frac{U}{2}}}\ \right]\rho \left (\sqrt{\omega^2-\frac{U^2}{4}}\right ),
\end{equation}
with $\rho(\epsilon)$ the noninteracting DOS at $U=0$ (and given by $\rho(\epsilon)=\exp(-\epsilon^2)/(t^*\sqrt{\pi})$ in the infinite-dimensional limit).
The average DOS is found by summing over the local DOS for each sublattice with weight $1/2$.
\begin{figure}
\centerline{\includegraphics[width=3.5in,clip=on]{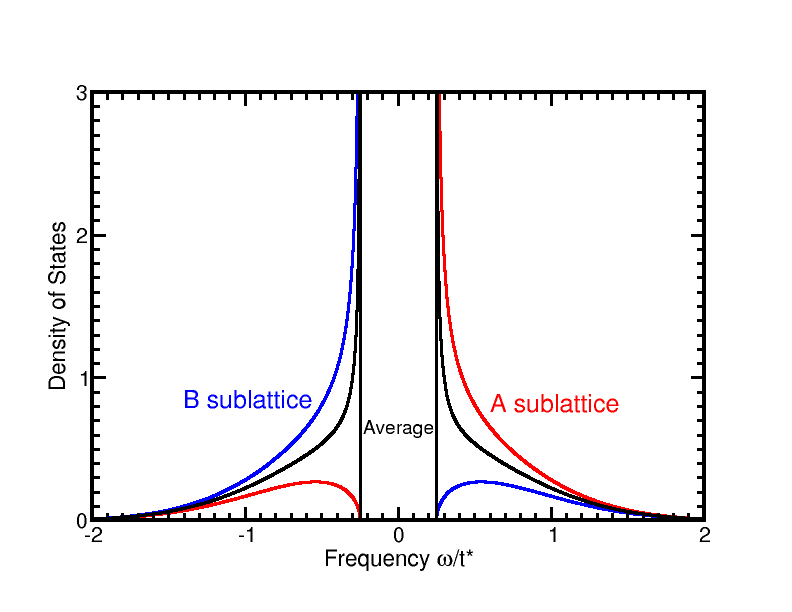}}
\caption{(Color online.) Equilibrium density of states for $U=0.5$ at half filling. The red curve is the local DOS on the $A$ sublattice, the blue curve on the $B$ sublattice and the black curve is the average local DOS.} \label{fig: dos_eq}
\end{figure}
Fig.~\ref{fig: dos_eq} shows the equilibrium DOS for $U=0.5$. Note that the band gap is equal to the onsite potential $U$, which shows the system is an insulator for nonzero $U$. The  local DOS on the $A$ sublattice (red line) has a divergence at $\omega=U/2$, while the local DOS on the $B$ sublattice (blue line) has a divergence at $\omega=-U/2$ (the black curve is the average local  DOS).

In equilibrium, the local  lesser Green's function, defined by
\begin{equation}
G_i^<(t)=i\langle c_i^\dagger(0)c_i^{}(t)\rangle,
\end{equation}
satisfies $G_i^<(\omega)=-2if(\omega){\rm Im}G_i^R(\omega)$, with $f(\omega)=1/[1+\exp(\beta\omega)]$ the Fermi-Dirac distribution function.  At $T=0$, $f(\omega)$ vanishes for positive frequency and is equal to one for negative frequency, so the local equilibrium lesser function at $T=0$  is simply given by the left hand side of Fig.~\ref{fig: dos_eq}.

\begin{figure}
\centerline{\includegraphics[width=3.5in,clip=on]{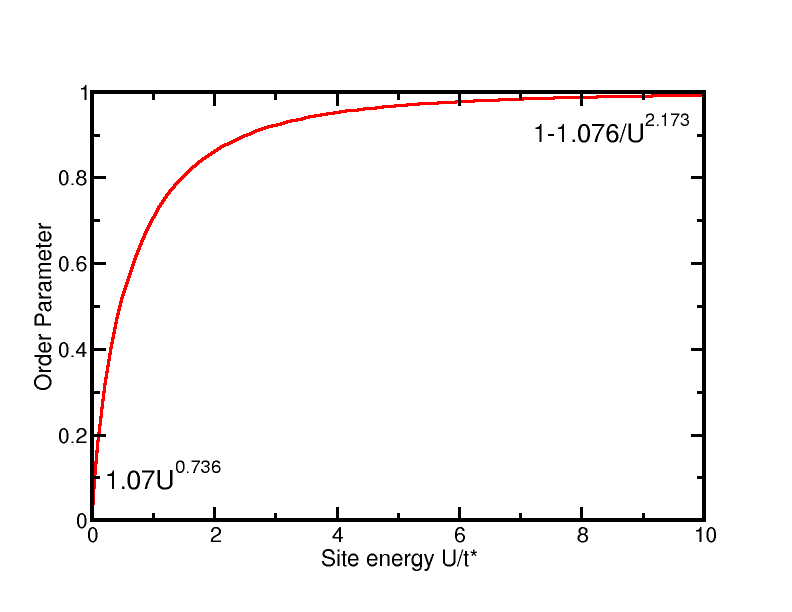}}
\caption{(Color online.) Equilibrium CDW order parameter as a function of $U$ at zero temperature. } \label{fig: order_eq}
\end{figure}

The lesser Green's function also gives us the local density of electrons on each sublattice
\begin{equation}
n_{A,B}={\rm Im}[G^{<}_{A,B}(t=0)]=\int_{-\infty}^{\infty} \frac{d\omega}{2\pi} {\rm Im}\left [ G_{A,B}^<(\omega)\right ].
\end{equation}
The equilibrium order parameter for the conduction electrons is just the difference between the electron number density on the $A$ and $B$ sublattice:
\begin{equation}
\Omega=\frac{n_B-n_A}{n_A+n_B}.
\end{equation}
Since there is a repulsive potential on the $A$ sublattice, there are always more electrons on the $B$ sublattice than on the $A$ sublattice in equilibrium and the order parameter is maximal for fixed $U$ when the temperature is equal to zero.
The equilibrium order parameter at zero temperature is plotted in Fig.~\ref{fig: order_eq} as a function of the onsite potential $U$. Formulas for empirical fits for small and large $U$ are also shown. 

\subsection{Nonequilibrium formalism}

In the case of nonequilibrium, the Hamiltonian becomes time dependent due to the presence of an electric field. We include this electric field via the Peierls' substitution\cite{Peierls_gauge}, which is a simplified semiclassical treatment of the electromagnetic field that is exact and nonperturbative. With the Peierls' substitution, the hopping matrix gains a time dependent phase factor\cite{Jauho} due to the field
\begin{equation}
t_{ij}\rightarrow t_{ij}(t)=t_{ij}\exp\left[-\frac{ie}{\hbar c}\int_{R_i}^{R_j}\textbf {A} (\textbf{r},\emph{t})\cdot d\textbf {r}\right].
\end{equation}
This result just follows from the phase a particle picks up when moving under the influence of a vector potential and is sufficient to describe the electric field when we work in a gauge where there is zero scalar potential and only a time-dependent vector potential. The electric field $\textbf{E} (\textbf{r},\emph{t})$ is found from the temporal derivative of the vector potential $\textbf{A}(\textbf{r}, \emph{t})$.
\begin{equation}
\textbf {E}( {\bf r},\emph{t})=-\frac{1}{c}\frac{\partial {\textbf{A} (\textbf{r},\emph{t})}}{\partial t}.
\end{equation}
We will further assume that the electric field is spatially uniform, even when it is time dependent, neglecting the time-dependent magnetic field required by Maxwell's equations since those effects are much smaller than the electric field effects. We remark that the case considered here is one of the few in which Peierls' inclusion of the vector potential can be done gauge-invariantly, yet still result in photon-assisted transitions across the band gap. As pointed out in Ref.~\onlinecite{foreman}, a Peierls' construction often has limitations in multiband systems. However, since the gap in this model results from a site-potential that couples electrons with the same symmetry, interband transitions can be driven by the field via the Peierls' substitution.

We choose the spatially uniform field to lie along the diagonal direction, so $\textbf {A}(t)=A(t)(1,1,\ldots,1$). The time-dependent bandstructure in momentum space for the $U=0$ case then becomes,
\begin{equation}
\varepsilon_k(t)=-\displaystyle\sum_{ ij}t_{ij}\exp [-i(\textbf{k}-\frac{e}{\hbar c}\textbf {A} (\emph{t}))\cdot (\textbf{R}_{i}-\textbf{R}_{j})].
\end{equation}
So the effect of the Peierls' substitution is to add a time-dependent shift to the momentum in the noninteracting electronic band structure:
\begin{equation}
\varepsilon_k(t)=-\lim_{d\rightarrow\infty}\frac{t^*}{\sqrt{d}}\displaystyle\sum_{l=1}^{d}\cos\left[a\left(k_l-\frac{e{A}(t)}{\hbar c}\right)\right].
\end{equation}
The time-dependent Hamiltonian in the Schr\"odinger picture then becomes
\begin{eqnarray}
&~&\mathcal{H}_S(t)= \\
&~&\displaystyle\sum_{k}\overgroup{\undergroup{\begin{array}{cc}\emph{c}_{k}^{\dagger}&\emph{c}_{k+Q}^{\dagger}\end{array}}}
\left(\begin{array}{cc}{\frac{U}{2}-\mu+\varepsilon_k(t)} & {\frac{U}{2}} \\ {\frac{U}{2}} & {\frac{U}{2}-\mu-\varepsilon_k(t)}\end{array}\right)
\left(\begin{array}{c}\emph{c}_{k}\\\emph{c}_{k+Q}\end{array}\right)\nonumber
\end{eqnarray}
in momentum space.
The time-dependent band structure $\varepsilon_k(t)$ can be expanded with the difference formula of the cosine (for the diagonal field) via
\begin{equation}
\varepsilon_k(t)= \cos\left(\frac{eaA(t)}{\hbar c}\right)\varepsilon_k +\sin\left(\frac{eaA(t)}{\hbar c}\right)\bar\varepsilon_k
\end{equation}
which depends on the band structure $\varepsilon(k)$
and the projection of the velocity along the field
\begin{equation}
\bar\varepsilon_k=-\lim_{d\rightarrow\infty}\frac{t^*}{\sqrt{d}}\displaystyle\sum_{l=1}^d\sin(ak_l).
\end{equation}
In the Heisenberg picture, we can write the equation of motion for the operators $c_k(t)$ and $c_{k+Q}(t)$, 
\begin{equation}
i\frac{dc_k(t)}{dt}=-\left[\mathcal{H}_H(t),c_k(t)\right] 
\end {equation}
and
\begin{equation}
i\frac{dc_{k+Q}(t)}{dt}=-\left[\mathcal{H}_H(t),c_{k+Q}(t)\right] ,
\end{equation}
where $\mathcal{H}_H(t)$ is the Heisenberg representation for the Hamiltonian.
If we substitute in the time-dependent Hamiltonian and evaluate the commutators,
we have
\begin{equation}
i\frac{dc_k(t)}{dt}=\sum_k \left [\left (\frac{U}{2}-\mu+\varepsilon_k(t) \right )c_k(t)+\frac{U}{2}c_{k+Q}(t)\right ]
\end{equation}
\begin{equation}
i\frac{dc_{k+Q}(t)}{dt}=\sum_k \left [ \frac{U}{2}c_{k}(t)+\left (\frac{U}{2}-\mu-\varepsilon_k(t) \right )c_{k+Q}(t)\right ].
\end{equation}
Then we have the time evolution for the annihilation operators satisfies
\begin{equation}
\left(\begin{array}{c}c_{k}(t)\\c_{k+Q}(t)\end{array}\right)=U(k,t,t_0)\left(\begin{array}{c}c_{k}(t_0)\\c_{k+Q}(t_0)\end{array}\right).
\end{equation}
The time-evolution operator $U_(k,t,t')$ is a time ordered product for each momentum
\begin{eqnarray}
&~&\emph U(k,t,t')=\\
&~&\emph T_t \exp\left[- i\int_{t'}^t d\bar{t}\left(\begin{array}{cc}{\frac{U}{2}-\mu+\varepsilon_k(\bar{t})} & {\frac{U}{2}} \\ {\frac{U}{2}} &{\frac{U}{2}-\mu-\varepsilon_k(\bar{t})} \end{array}\right)\right].\nonumber
\end{eqnarray}
Since there are time-dependent terms inside the exponential, we must numerically calculate the time evolution $U(k,t,t')$ by employing the Trotter formula:
\begin{equation}
U(k,t,t')=U(k,t,t-\Delta t)U(k,t-\Delta t,t-2\Delta t)\cdots U(k,t'+\Delta t,t').
\end{equation}
For a small time step $\Delta t$ at time $t$, we have
\begin{eqnarray}
&~&U(k,t,t-\Delta t)=\\
&~&\exp\left[-i\Delta t\left(\begin{array}{cc}{\frac{U}{2}-\mu+\varepsilon_k(t-\Delta t/2)} & {\frac{U}{2}} \\ {\frac{U}{2}} &{\frac{U}{2}-\mu-\varepsilon_k(t-\Delta t/2)}\end{array}\right)\right].\nonumber
\end{eqnarray}
This exponential can be exactly found since it is a $2\times 2$ matrix, and we show the result for the case
of interest of half-filling, where $\mu=U/2$
\begin{eqnarray}
&~&U(k,t,t-\Delta t)=\cos \left (\Delta t\sqrt{\varepsilon_k^2\left (t-\frac{\Delta t}{2}\right )+\frac{U^2}{4}}\right )\textbf{I} \nonumber\\ &~&-i \left(\begin{array}{cc}{\varepsilon_k\left (t-\frac{\Delta t}{2}\right )} & {\frac{U}{2}} \\ {\frac{U}{2}} &{-\varepsilon_k \left ( t-\frac{\Delta t}{2}\right )}\end{array}\right)\nonumber\\
&~&\times\frac{\sin \left (\Delta t{\sqrt{\varepsilon_k^2\left (t-\frac{\Delta t}{2}\right)+\frac{U^2}{4}}}\right )}{ \sqrt{\varepsilon^2_k\left (t-\frac{\Delta t}{2}\right )+\frac{U^2}{4}}}.
\end{eqnarray}
In our calculations, we must start from a finite minimum time instead of $t_{min}\rightarrow -\infty$, so we calculate the time-evolution operator from a minimal time $t_0$: $U(k,t,t_0)$.
For each $k$, we find the two-time evolution operator from the identity
\begin{equation}
U(k,t,t')=U(k,t,t_0)U^{\dagger}(k,t_0,t').
\label{eq: evolution_identity}
\end{equation}
Once the time evolution at each time pair is found, we then calculate the nonequilibrium Green's functions to obtain the physical properties of the system.  Therefore, one can see that the exact solution in the nonequilibrium case is much more complex than the equilibrium solution due to the change in time of the instantaneous eigenbasis; this then requires significant numerical resources to properly solve the problem.

\begin{figure}
\centerline{\includegraphics[width=3.5in,clip=on]{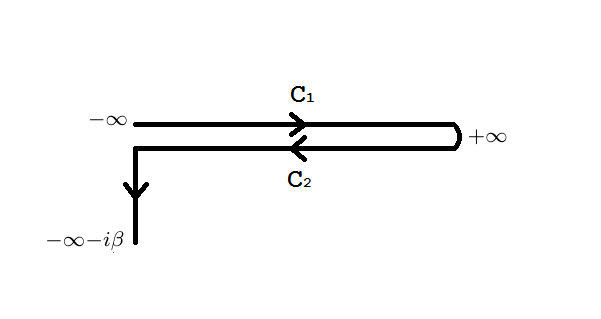}}
\caption{Kadanoff-Baym-Keldysh contour. The contour evolves from a minimal time to a maximal one, then evolves backwards to the minimal one, and finally evolves parallel to the negative imaginary time axis out to a distance given by $\beta=1/T$.} \label{fig: contour}
\end{figure}

The contour-ordered single-particle Green's function along the Kadanoff-Baym-Keldysh contour (see Fig.~\ref{fig: contour}) is defined as
\begin{equation}
\begin{array}{c}
{G^c_{ij}(t,t')=-i\langle T_c{c_i^{}(t)c^{\dagger}_j(t')\rangle}}\\ {=-i\theta_c (t,t')\langle c_i^{}(t)c^{\dagger}_j(t')\rangle+i\theta_c (t',t)\langle c^{\dagger}_j(t')c_i^{}(t)\rangle },
\end{array}
\end{equation}
where $\theta_c$ is the generalization of the unit step function to the contour and vanishes if $t$ is before $t'$ on the contour and is equal to one if $t$ is after $t'$ on the contour.
The contour-ordering operator $T_c$  orders the two operators $a(t)$ and $b(t')$ according to which is first along the contour (later along the contour on the left), and we have
\begin{equation}
T_c \{a(t)b(t')\}=\left \{ \begin{array}{c}{a(t)b(t') \ \ \ t'\ \textrm{before\ } t \textrm{\ on contour}}\\ {- b(t')a(t) \ \ \ t'\ \textrm{after\ }t \textrm{\ on contour}}\end{array}\right .
\end{equation}
The minus sign arises because $a(t)$ and $b(t')$ are fermionic operators. 

The retarded and lesser Green's functions continue to be defined by
\begin {equation}
G^{R}_{ij}(t,t')=-i\theta (t-t')\langle \{c_i^{}(t),c^{\dagger}_j(t')\}_+\rangle
\end{equation}
and
\begin {equation}
G^{<}_{ij}(t,t')=i\langle c^{\dagger}_j(t')c_i^{}(t)\rangle
\label{eq: lesser_noneq_def}
\end{equation}
where the operators are in the Heisenberg picture with respect to the time-dependent Hamiltonian, but they now depend on two times instead of just on the time difference.

Assuming the system starts at a time $t_0$ well before the field is turned on, so the system is in equilibrium, we can introduce the evolution operators for the time-dependent creation and annihilation operators and evaluate the retarded Green's function as follows
\begin{eqnarray}
G^R_{ii} (t,t')&=& -i \theta (t-t')\\
&\times&\langle \{c_k(t_0)U_{11}(k,t,t_0)+c_{k+Q}(t_0)U_{12}(k,t,t_0)\nonumber\\
&\pm& c_k(t_0)U_{21}(k,t,t_0)\pm c_{k+Q}(t_0)U_{22}(k,t,t_0),\nonumber\\
&~&c^{\dagger}_k(t_0)U^{\dagger}_{11}(k,t_0,t')+c^{\dagger}_{k+Q}(t_0)U^{\dagger}_{21}(k,t_0,t')\nonumber\\
&\pm& c^{\dagger}_k(t_0)U^{\dagger}_{12}(k,t_0,t')\pm c^{\dagger}_{k+Q}(t_0)U^{\dagger}_{22}(k,t_0,t')\}_+\rangle .\nonumber
\end{eqnarray}
The symbols $U_{ab}(k,t,t')$ and $U^{\dagger}_{ab}(k,t,t')$ represent the elements at row $a$ and column $b$ of the evolution matrices $U(k,t,t')$ and $U^{\dagger}(k,t,t')$. 
Evaluating the anticommutator, and using the identity of the evolution matrices in Eq.~(\ref{eq: evolution_identity}), then shows that the local retarded Green's function is just a function of the time-evolution between the two times $t$ and $t'$:
\begin{eqnarray}
G^R_{ii} (t,t')&=&-i \theta (t-t')\displaystyle\sum_k\{U_{11}(k,t,t')+U_{22}(k,t,t')\nonumber\\
&\pm& U_{12}(k,t,t')\pm U_{21}(k,t,t')\}.
\label{eq: g_ret_local}
\end{eqnarray}
The plus sign is for $i\in A$ sublattice, while the minus sign is for $i \in B$ sublattice because of the phase shift $\exp(\pm i\vec{Q}\cdot R_A)=1$ and $\exp(\pm i\vec{Q}\cdot R_B)=-1$ . The retarded Green's function determines the character of the quantum states of the system and does not depend on the history, which is why it depends only on the times between $t$ and $t'$. Note that this is not the same as saying it depends on the time difference $t-t'$ since the retarded Green's function does change with average time and the evolution operator is a complicated function of $t$ and $t'$ (see below).

Now we can introduce Wigner's average and relative time coordinates\cite{Wigner} which are defined via,
\begin{equation}
t_{rel}=t-t', \ \ t_{ave}=\frac{t+t'}{2}.
\end{equation}
The Fourier transform of the local retarded Green's function with respect to the relative time for a fixed average time
\begin{equation}
G^R_{ii} (\omega,t_{ave})=\int dt_{rel} e^{i\omega t_{rel}}G^R_{i}(t_{rel},t_{ave})
\end{equation}
yields the transient nonequilibrium local DOS $A_{ii}(\omega,t_{ave})=-{\rm Im}G^R_{ii}(\omega,t_{ave})/\pi$.

This local DOS on each sublattice satisfies a number of exact sum-rule relations even in nonequilibrium\cite{sumrules} which are proved in  Appendix A. This includes the zeroth through third moment.

In addition to the retarded Green's function, we also need the lesser Green's function to calculate the current, the kinetic and potential energy, the CDW order parameter, and the fillings in different bands as functions of time.
With the Fourier transform of the definition of the lesser Green's function in Eq.~(\ref{eq: lesser_noneq_def}), we can write the momentum-dependent lesser Green's functions as
\begin{eqnarray}
G^{<}_{11}(k,t,t')&=&i\langle c_k^{\dagger}(t')c_k(t)\rangle\\ 
&=&i U_{11}^{\dagger}(k,t_0,t')U_{11}(k,t,t_0)\langle c_k^{\dagger}(t_0)c_k(t_0)\rangle\nonumber\\ &+&iU_{11}^{\dagger}(k,t_0,t')U_{12}(k,t,t_0)\langle c_k^{\dagger}(t_0)c_{k+Q}(t_0)\rangle \nonumber\\ &+&iU_{21}^{\dagger}(k,t_0,t')U_{11}(k,t,t_0)\langle c_{k+Q}^{\dagger}(t_0) c_k(t_0)\rangle\nonumber\\
&+&iU_{21}^{\dagger}(k,t_0,t')U_{12}(k,t,t_0) \langle c_{k+Q}^{\dagger}(t_0)c_{k+Q}(t_0) \rangle ,
\nonumber
\end{eqnarray}
\begin{eqnarray}
G^{<}_{12}(k,t,t')&=&i\langle c_{k+Q}^{\dagger}(t')c_k(t)\rangle\\ 
&=& iU_{11}^{\dagger}(k,t_0,t')U_{21}(k,t,t_0)\langle c_k^{\dagger}(t_0)c_k(t_0)\rangle\nonumber\\ &+&iU_{11}^{\dagger}(k,t_0,t')U_{22}(k,t,t_0)\langle c_k^{\dagger}(t_0)c_{k+Q}(t_0)\rangle \nonumber\\ 
&+&iU_{21}^{\dagger}(k,t_0,t')U_{21}(k,t,t_0)\langle c_{k+Q}^{\dagger}(t_0) c_k(t_0)\rangle\nonumber\\
&+&iU_{21}^{\dagger}(k,t_0,t')U_{22}(k,t,t_0) \langle c_{k+Q}^{\dagger}(t_0)c_{k+Q}(t_0) \rangle ,
\nonumber
\end{eqnarray}
\begin{eqnarray}
G^{<}_{21}(k,t,t')&=&i\langle c_k^{\dagger}(t')c_{k+Q}(t)\rangle\\ 
&=& iU_{12}^{\dagger}(k,t_0,t')U_{11}(k,t,t_0)\langle c_k^{\dagger}(t_0)c_k(t_0)\rangle\nonumber\\ &+&iU_{22}^{\dagger}(k,t_0,t')U_{12}(k,t,t_0)\langle c_k^{\dagger}(t_0)c_{k+Q}(t_0)\rangle \nonumber\\ 
&+&iU_{12}^{\dagger}(k,t_0,t')U_{11}(k,t,t_0)\langle c_{k+Q}^{\dagger}(t_0) c_k(t_0)\rangle\nonumber\\
&+&iU_{22}^{\dagger}(k,t_0,t')U_{12}(k,t,t_0) \langle c_{k+Q}^{\dagger}(t_0)c_{k+Q}(t_0) \rangle ,
\nonumber
\end{eqnarray}
and
\begin{eqnarray}
G^{<}_{22}(k,t,t')&=&i\langle c_{k+Q}^{\dagger}(t')c_{k+Q}(t)\rangle\\ 
&=& iU_{12}^{\dagger}(k,t_0,t')U_{21}(k,t,t_0)\langle c_k^{\dagger}(t_0)c_k(t_0)\rangle\nonumber\\ &+&iU_{22}^{\dagger}(k,t_0,t')U_{22}(k,t,t_0)\langle c_k^{\dagger}(t_0)c_{k+Q}(t_0)\rangle \nonumber\\ 
&+&iU_{12}^{\dagger}(k,t_0,t')U_{21}(k,t,t_0)\langle c_{k+Q}^{\dagger}(t_0) c_k(t_0)\rangle\nonumber\\
&+&iU_{22}^{\dagger}(k,t_0,t')U_{22}(k,t,t_0) \langle c_{k+Q}^{\dagger}(t_0)c_{k+Q}(t_0) \rangle \nonumber .
\end{eqnarray}
Note how these results do not simplify to depend only on the times between $t$ and $t'$, but instead depend on all times. In addition, these results depend upon the initial occupancies of the different momentum states, which are determined by the initial conditions at $t=t_0$ and follow from the equilibrium analysis described above. 
In particular, we start the system in equilibrium at $T=0$ which corresponds to a filled lower band $\langle c_{k-}^\dagger c_{k-}^{}\rangle=1$ and an empty upper band $\langle c_{k+}^\dagger c_{k+}^{}\rangle=0$.  Converting from the eigenfunction basis to the $k$ and $k+Q$ basis then shows that we must take
$\langle c_k^\dagger c_k^{}\rangle=\beta_k^2$, $\langle c_{k+Q}^\dagger c_{k+Q}^{}\rangle=\alpha_k^2$, and
$\langle c_k^\dagger c_{k+Q}^{}\rangle = \langle c_{k+Q}^\dagger c_k^{}\rangle=-\alpha_k\beta_k$ as the inital expectation values for the occupancies. Here, the $\alpha_k$ and $\beta_k$ are the equilibrium values in Eqs.~(\ref{eq: alpha_eq}) and (\ref{eq: beta_eq}) since the system starts in equilibrium before the field is turned on.

The time-dependent CDW order parameter then follows in a simple fashion
\begin{eqnarray}
\Omega(t)&=&\frac{n_B(t)-n_A(t)}{n_B(t)+n_A(t)}\\
&=&
-\frac{\sum_{k:\varepsilon_k\le 0}[G_{12}^<(k,t,t)+G_{21}^<(k,t,t)]}{\sum_{k:\varepsilon_k\le 0}
[G_{11}^<(k,t,t)+G_{22}^<(k,t,t)]}.\nonumber
\end{eqnarray}

\subsection{Gauge invariance}

Physically measurable properties are gauge invariant, and depend only upon the fields, not the scalar and vector potentials. While local quantities are always gauge invariant, quantities that depend upon momentum do depend on the gauge, and we need to make a transformation from the Green's functions in a particular gauge to the gauge-invariant Green's functions in order to determine those quantities~\cite{Jauho,freericks_book}
We let ${\bf G}(k,t_{rel},t_{ave})$ denote the $2\times 2$ matrix for the Green's function in the gauge with momentum $k$, relative time $t_{rel}$ and average time $t_{ave}$. A superscript of $R$ or $<$ will denote the retarded or lesser Green's function. In the gauge, the reduced Brillouin zone is defined by $\epsilon_k<0$.  When we go to the gauge-invariant Green's function [denoted by $\tilde {\bf G}(k,t_{rel},t_{ave})={\bf G}(k(t_{rel},t_{ave}),t_{rel},t_{ave})$], we simply make the transformation 
\begin{equation}
k\rightarrow k(t_{rel},t_{ave})= k+\int_{-\frac{1}{2}}^{\frac{1}{2}}d\lambda A(t_{ave}+\lambda t_{rel})
\end{equation}
which means the reduced Brillouin zone for the gauge-invariant Green's function satisfies
$\varepsilon_{k(t_{rel},t_{ave})}\le 0$. This is an added complication that one has to deal with in a gauge-invariant formulation for a system with reduced spatial symmetry, because if we had the full Brillouin zone, it would be identical for the gauge-invariant case and the case in a gauge.  Instead, the reduced Brillouin zone becomes time-dependent for the gauge-invariant formalism.

Our strategy for calculating the current is to use the gauge-invariant Green's function. We will first determine the linear-response formula for the current, and then generalize it to the nonlinear-response case.  This is simple to do in a gauge-invariant formalism because the formula is independent of the vector potential.  The current density operator is defined as the commutator of the Hamiltonian and the charge polarization 
\begin{equation}
\textbf{j}(t)=i[H(t),\displaystyle\sum\limits_{j} {\bf R}_j c_j^{\dagger}(t)c_j(t)],
\end{equation}
which becomes
\begin{equation}
\textbf{j}(t)=-i\sum_{i,\delta}t_{i,i+\delta}(t)\vec{\delta} c_i^{\dagger}(t) c_{i+\delta}(t)
\end{equation}
where $\delta$ denotes a nearest neighbor translation vector from site $i$. In linear response, we ignore the time-dependence of the hopping.
Next, we write the linear-response current by summing only over the $A$ sublattice because the field is uniform and we have current conservation in the system. With the field along the diagonal, the current along each spatial component can be written as, 
\begin{equation}
j_{\alpha}(t)=-it\sum_{i\in A,\delta}\delta_\alpha [c_i^{\dagger}(t) c_{i+\delta}(t)+ c_{i+\delta}^{\dagger}(t) c_{i+2\delta}(t)],
\end{equation}
with the second term coming from the $B$ sublattice contributions.
We now convert to a momentum-space representation, which gives
\begin{equation}
j_{\alpha}(t)=\displaystyle\sum\limits_{k:\varepsilon_k\le 0}\boldsymbol{\nabla}_{k_\alpha}\varepsilon_k [ c_k^{\dagger}(t)c_{k}(t)-c_{k+Q}^{\dagger}(t)c_{k+Q}(t)]
\end{equation}
and is easily recognizable as the correct formula for the linear-response current.

To generalize this formula to the nonequilibrium case, we evaluate the expectation value of the current operator by using the gauge-invariant Green's function
\begin{equation}
\langle j_{\alpha}(t)\rangle=\displaystyle\sum\limits_{k:\varepsilon_{k+A(t)}\le 0}\boldsymbol{\nabla}_{k_\alpha}\varepsilon_k[ \tilde{G}^<_{11}(k,t,t)-\tilde{G}^<_{22}(k,t,t)].
\end{equation}
This formula has been written for the nonequilibrium case since the linear-response form of the current has no dependence on the vector potential, and hence is the correct form to generalize to nonequilibrium. 
Using the fact that $\tilde {\bf G}^<(k,t_{rel}=0,t_{ave})={\bf G}^<(k+A(t_{ave}),t_{rel}=0,t_{ave})$ then shows that
\begin{eqnarray}
\langle j_{\alpha}(t)\rangle&=&\displaystyle\sum\limits_{k:\varepsilon_{k+A(t)}\le 0}\boldsymbol{\nabla}_{k_\alpha}\varepsilon_k [ {G}^<_{11}(k+A(t),t,t)\nonumber\\
&-&{G}^<_{22}(k+A(t),t,t)]
\end{eqnarray}
where we changed to using the $t$, $t'$ representation for the Green's function.  Now, we simply shift $k\rightarrow k-A(t)$ to get our final formula for the current
\begin{equation}
\langle j_{\alpha}(t)\rangle=\displaystyle\sum\limits_{k:\varepsilon_{k}\le 0}\boldsymbol{\nabla}_{k_\alpha}\varepsilon_{k-A(t)} [ {G}^<_{11}(k,t,t)-{G}^<_{22}(k,t,t)].
\label{eq: current}
\end{equation}
One can also derive the current in the more traditional way by working with the formulation entirely in the Hamiltonian gauge, and the result is identical.

\section{Bloch oscillation in a CDW system}
In a pure material composed of electrons interacting with the periodic lattice potential, the system does not satisfy Ohm's law.  Instead, if a field is applied to the material, it generates an oscillating current called a Bloch oscillation\cite{Bloch,Zener}. Ohm's law is recovered when one introduces scattering into the system and the field is small enough to be in the linear-response regime. As surprising as this result might be, it is even more surprising in that it is quite difficult to observe in real materials, because the scattering time, for even the most pure materials, is too short for the system to show the Bloch oscillation. It has been  observed in semiconductor superlattices \cite{Blochinsemiconductor,Blochinsemiconductor1,Blochinsemiconductor2}, in cold atoms in an optical lattice\cite{Blochincoldatoms1,Blochincoldatoms} and also in ultra-small Josephson junctions \cite{BlochinJosephsen}.

If we go back to the theory we developed in Section II, and set $U=0$ to recover the single-band model, then one immediately can solve for the evolution operators, because the $2\times 2$ matrix becomes diagonal.  If we work in the full Brillouin zone, then we find that the current satisfies
\begin{eqnarray}
\langle j_\alpha(t)\rangle &=& \int d\varepsilon \int d\bar\varepsilon \rho(\varepsilon)\rho(\bar\varepsilon)
f(\varepsilon) \\\
&\times&[-\bar\varepsilon \cos(A(t))+\varepsilon\sin(A(t))]\nonumber
\end{eqnarray}
which is proportional to $\sin(Et)$, with a temperature-dependent amplitude. The current oscillates about zero, so the energy added to the system also oscillates about zero, and after each Bloch period, given by $2\pi/E$, the system returns to the original state it was in before the field was put on.  This is a well-known property of Bloch oscillations in a noninteracting single-band model. When scattering is added into the system, the current oscillates about a net positive value, so that a nonzero amount of heat is always added to the system. 

Our goal here, is to study how this situation changes when we have a two-band model that arises from the presence of CDW order.  Even though the system is noninteracting, we no longer have any guarantee that the system can return to its initial state after a Bloch period. Since the excitation of electrons across the gap requires quantum-mechanical tunneling across the gap, it is not obvious that the electrons excited across the gap can all be de-excited at any specific time.  Indeed, we will find out that this does not occur in general. 
 
\begin{figure} 
\centerline{\includegraphics[width=95mm,clip=on]{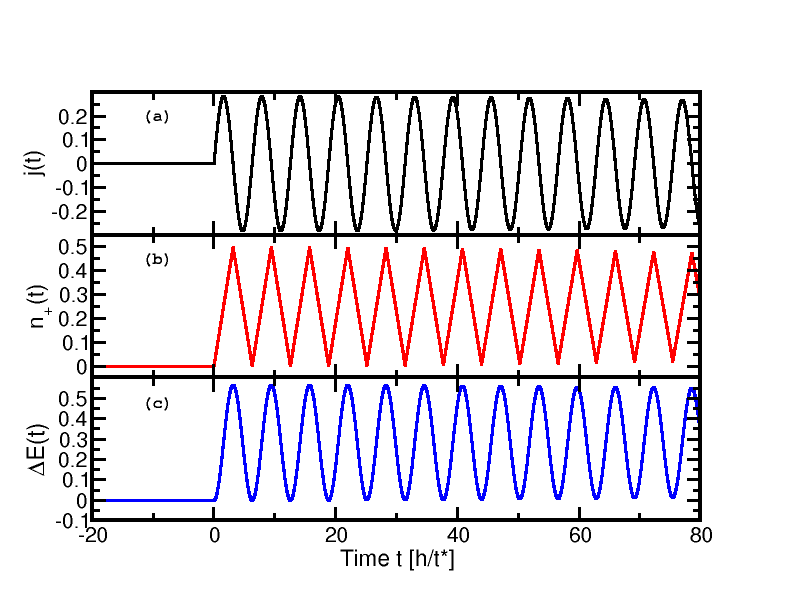}}
\caption{(Color online.) Current, upper band occupancy and change in energy for a $U=0.01$ CDW system placed in a DC field with $E(t)=\theta(t)$. Panel (a) shows the current, which is close to the known sine wave that occurs at $U=0$ with a frequency equal to the amplitude of the DC field ($\omega=E_0=1$). Panel (b) shows the upper band electron occupancy as a function of time. Panel (c) shows the total energy shift from the ground state energy  as a function of time.} \label{fig: curr_u=0.01}
\end{figure}

Based on the theory developed in the Sec. II, we can calculate the current with a DC field using the nonequilibrium Green's function technique. The procedure is to discretize the contour, calculate the relevant evolution operators for each momentum point, and use them to construct the retarded and lesser Green's functions as functions of momentum and time. From these Green's functions all relevant physical quantities can be calculated. We turn on a spatially uniform DC electric field of amplitude $E_0$ abruptly at time $t=0$, and calculate the subsequent evolution of the system.  To verify the formalism, we examine the case with $U=0.01$ and $E_0=1$, where the DC field is much larger than the gap size, so the system should have similar behavior to the single-band model and illustrate similar Bloch oscillations.
Fig.~\ref{fig: curr_u=0.01} shows the current, the upper band electron occupancy and the total energy for this case. The current oscillates with the Bloch frequency and has a small reduction of the amplitude over time due to the dephasing in the system (because there are two bands since $U\ne 0$). Note how the current appears to oscillate about zero, but because of the decaying amplitude, there is a net current, when integrated over time, and hence there is a small net transfer of energy into the system.
The upper band electron occupancy has a nearly sawtooth nature to it as electrons are excited and then de-excited from the upper band. The electrons move between the lower and upper bands almost as if they don't feel the presence of the gap. The increase of energy as a function of time appears to be quite similar to the single-band case, but it doesn't actually go to zero at the Bloch period; instead it has a small residual energy gain there.

\begin{figure} 
\centerline{\includegraphics[width=100mm,clip=on]{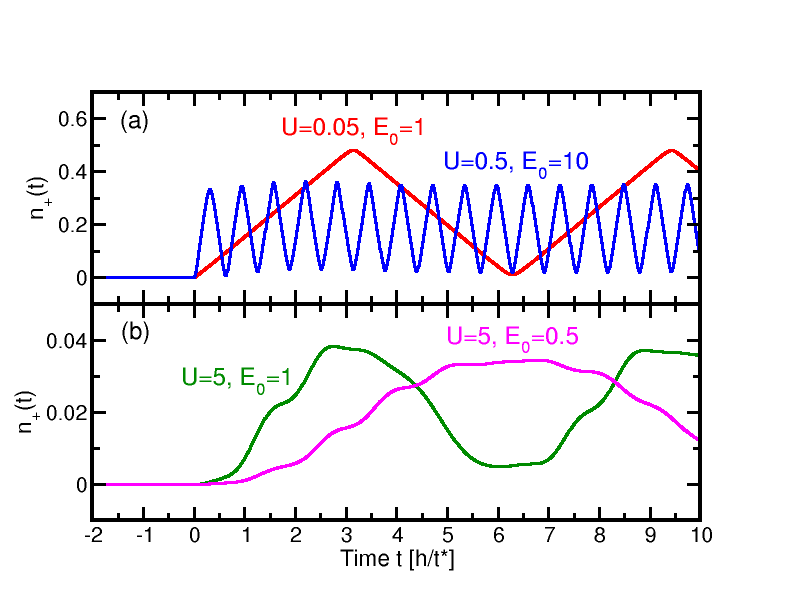}}
\caption{(Color online.) Dielectric breakdown of the CDW insulator for large electric fields. The upper band filling  is plotted for different $U$ and $E_0$ values. When $E_0\gg U$, as shown in panel (a), the upper band electrons are excited as if there is no gap, although we never fully fill the upper band, which occurs when $n_+=0.5$.  Here, we show two cases with the same ratio of $U/E_0$: $U=0.05$ and $E_0=1$ (red curve, long period) and $U=0.5, E_0=10$ (blue curve, short period). Note how the total amount excited depends on more than just the ratio of $U/E_0$. When $U\gg E_0$ [panel (b), note change in vertical scale], much fewer electrons are pumped to the upper band. Here we show a large gap $U=5$ and smaller fields $E_0=1$ (violet, short period) and $E_0=0.5$ (magenta, long period). Note how the maximum amplitude decreases as the gap increases by comparing panel (a) to panel (b). } \label{fig: dielectric_breakdown}
\end{figure}

We illustrate the behavior of dielectric breakdown, where current is induced across the gap due to a large electric field, in Fig.~\ref{fig: dielectric_breakdown}.  When the field magnitude is much smaller than the gap, field assisted tunneling across the gap is a rare event that is difficult to achieve because the tunneling process sees a large barrier, while the opposite case, where the field magnitude is much larger than the gap, readily has field-assisted tunneling occur. The bottom panel shows the former case, while the top panel shows the latter. Because quantum tunneling always occurs, no matter how small the field amplitude is, there is no sharp distinction between the case where Bloch oscillations readily occur and where they are suppressed.  It is instead a crossover.  But we expect the crossover to occur close to the region where $E_0=U$, since the gap is always equal to $U$ in equilibrium.

\begin{figure} 
\centerline{\includegraphics[width=100mm,clip=on]{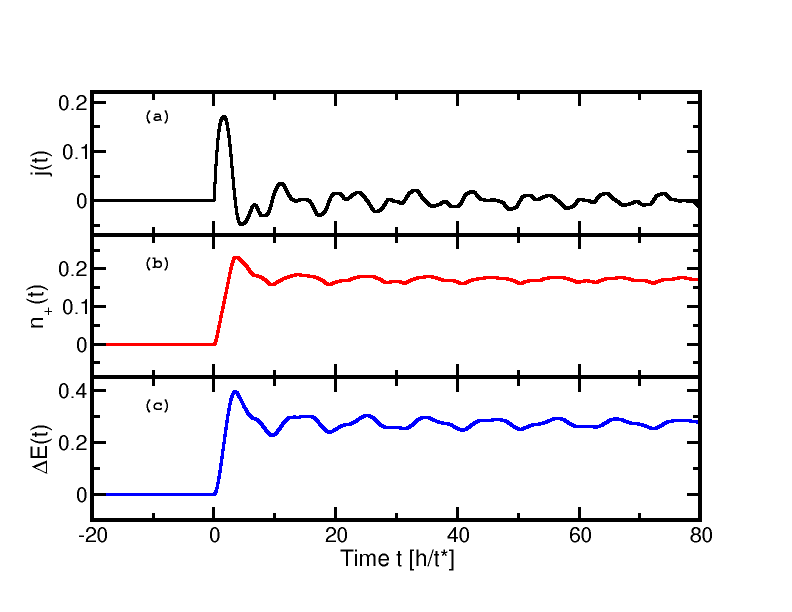}}
\caption{(Color online.) Results for $U=1$ in a DC field $E(t)=\theta(t)$ ($E_0=1$). Panel (a) shows the current, which has an initial transient response that settles down to a ``steady state'' and shows complex oscillations that are not given solely by the Bloch oscillation frequency. Panel (b) shows the upper band filling as a function of time as the system is driven by the electric field. Panel (c) shows the total energy as a function of time.} \label{fig: u=1}
\end{figure}

We next look at what happens at the crossover where the dielectric breakdown occurs $U=E_0$ in Fig.~\ref{fig: u=1} with $U=E_0=1$.
In this case, the current is initially driven to a relatively large amplitude and then stabilizes and shows complex oscillations. The transient response settles down rapidly to a more steady state behavior.  This occurs even though there is no scattering, and hence must come from dephasing effects. About 40$\%$ of the electrons are driven to the upper band as the initial field is switched on. The upper band electron number transiently oscillates, but maintains a level around 40$\%$ of the total number of electrons. In this quasi-steady state, the current oscillates around zero, so there is no net increase in the energy transfered to the system after we absorbed the initial energy that occurred during the transient response. It is clear in this case, that the system will not return back to the state it initially was in after the Bloch period. This is one of the main differences that occurs for the noninteracting model in a two-band system versus a single-band system.

\begin{figure} 
\centerline{\includegraphics[width=96mm,clip=on]{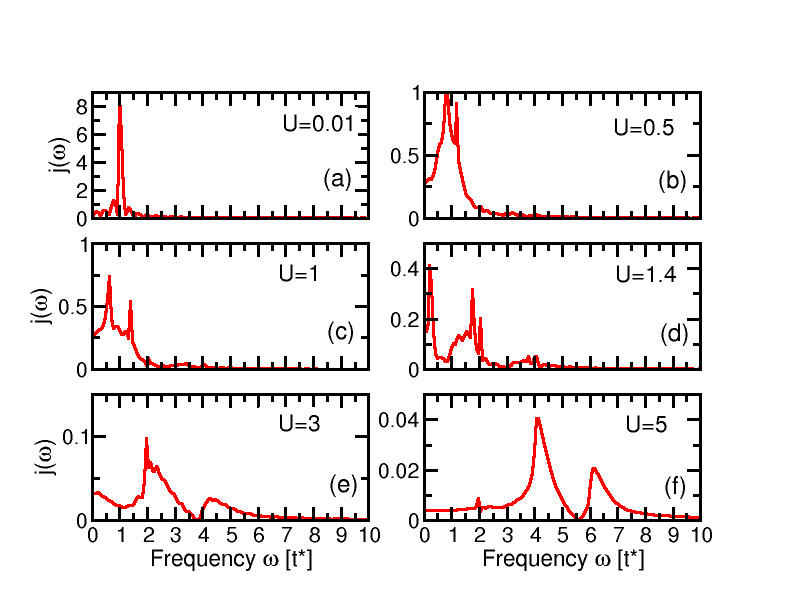}}
\caption{(Color online.) Fourier transform of the current response $j(\omega)$ to a DC field $E=\theta(t)$ ($E_0=1$) for different $U$  values.} \label{fig: ft_current}
\end{figure}

The effect of scattering on Bloch oscillations driven by a DC field (in the high-temperature disordered phase) has been studied in the Falicov-Kimball model\cite{jkf_noneq1,jkf_noneq2} and in the Hubbard model\cite{eckstein_hubbard} as has photoexcitation\cite{brian1,brian2,eckstein2}.  There are two important energy scales in the problem.  The first is the field amplitude $E_0$ and the second is the scattering strength, which is commonly denoted by $U_{int}$.  In the case of weak scattering, two scenarios are possible for the evolution of the local density of states: (i) the delta function peaks of the Wannier-Stark ladder (a series of delta function peaks in the local DOS separated by the Bloch frequency $\omega_B=E$) broaden due to scattering but remain near the Bloch frequencies or (ii) the delta function peaks split due to $U_{int}$ and broaden, but each Wannier-Stark miniband has a Mott-like transition. In cases where the interaction is strong enough to drive the system into an insulating phase, more complex behavior can occur.  For the Falicov-Kimball model, one still sees remnants of the Wannier-Stark physics\cite{freericks_book,aoki}, while they seem to be more strongly suppressed in the Hubbard model\cite{joura_hubbard,eckstein_hubbard_strong,karlis}. 

One way to try to identify the important energy scales is to look at the oscillation frequencies that arise in the current that is driven by the field.
To do this, we use a discrete Fourier transform of the current versus time traces, over a range of time where the transient effects have died off and the system is in a quasi-steady state (which is actually a nearly periodically varying state with a period given by the Bloch period); the time interval we used is $40\le t\le 80$. These results are plotted in Fig.~\ref{fig: ft_current}. We find that when $U$ is much larger than the field amplitude $E_0$, the current oscillates at frequencies $U\pm E_0$ as shown in the $U=5$ case. When $U<2E_0$, peaks appear around $E_0\pm U/2$. For intermediate values, these two effects mix as indicated with the $U=3$ case. Hence in this two band system, when the DC field is applied, the band gap $U$ and the electric field amplitude are the two main factors that affect the electron oscillations and these two factors interact with each other to produce new oscillation frequencies, that are not simply what might have been predicted by just looking at the two energy scales.

\begin{figure} 
\centerline{\includegraphics[width=100mm,clip=on]{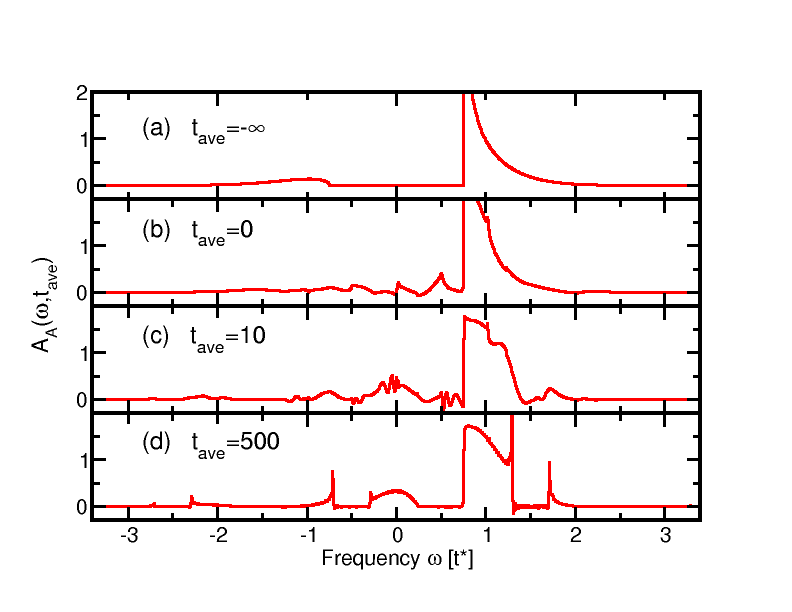}}
\caption{(Color online.) Local DOS for $U=1.5$ at different average times with $E_0=1$ and on the $A$ sublattice. The first panel is the equilibrium result ($t_{ave}\rightarrow -\infty$). Panel (b) has $t_{ave}=0$ and corresponds to when the field is turned on. Panel (c) has $t_{ave}=10$ and hence is in the transient response regime. Panel (d) has $t_{ave}=500$ and is approaching the steady state.} \label{fig: ldos_u=1.5}
\end{figure}

As we hinted at above, it is also interesting to examine the nonequilibrium local DOS. We show this DOS in Fig.~\ref{fig: ldos_u=1.5} for a number of different average times. Even though one might have expected the DOS to instantly switch between the equilibrium and the nonequilibrium results as the field is turned on, since the DOS measures the quantum mechanical states of the system, they must evolve from one result to the other continuously and this occurs slowly in this case because the equilibrium DOS has long tails in the time domain due to the inverse square root singularity at the upper or lower band edge. When we construct the retarded Green's function to describe these two systems and fix the average time, then for some relative times, we have one time before the field being turned on and one after.  This ``mixed'' Green's function interpolates between the equilibrium and nonequilibrium steady state DOS. If it also has long tails in time, then the evolution from equilibrium to nonequilibrium can be slow. 

One should note that due to the long tail in time for the equilibrium Green's function, it becomes difficult to think of the local DOS as being defined at a specific average time only.  The DOS involves the Fourier transform with respect to relative time, so it is defined in terms of two times, and one can have one of the times being long after the field is turned on, even if the average time is before the field is turned on. This is illustrated dramatically in panel (b), which has an average time equal to the time the field is turned on, but already shows significant deviations from the equilibrium DOS in panel (a). In particular, the gap is no longer well defined, and there is significant subgap structure. In the transient regime of panel (c), we see this evolve even further, and we also see the DOS go negative for some frequencies. This is not an artifact of a truncation of the Fourier transform, but commonly occurs for transient DOS in nonequilibrium calculations since there is no Lehmann representation for the transient DOS that allows us to prove nonnegativity of the DOS. Finally, in panel (d) we see the emergence of the steady state DOS, with its new gap structure occuring at the half-odd integers (except for the missing subband at $\omega=-1.5$); the separation of the gaps appears to be governed primarily by $E_0$ here. This result is similar to what one might expect due to Wannier-Stark physics, where the delta function peaks are split due to the CDW gap, but what is surprising here is the center of the CDW gap is occurring at half-odd integers here rather than at integers which is what the Wannier-Stark picture would predict.

\begin{figure} 
\centerline{\includegraphics[width=95mm,clip=on]{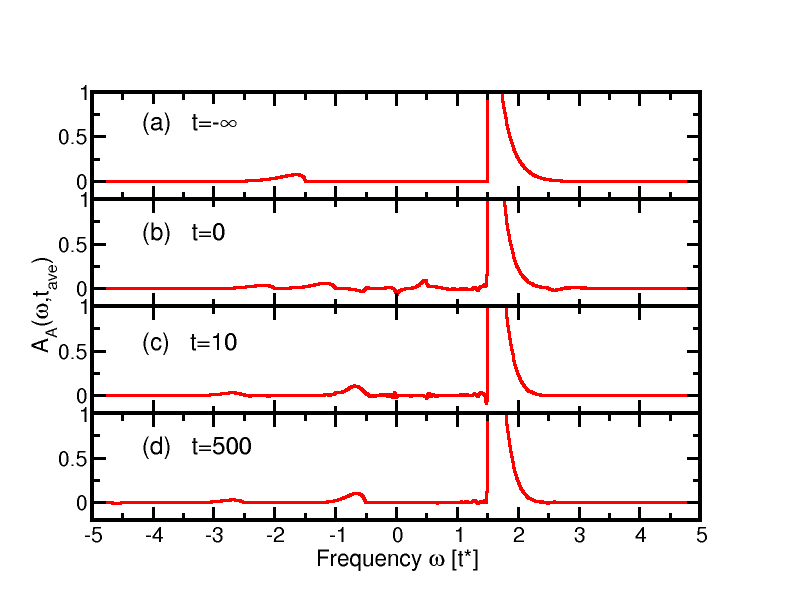}}
\caption{(Color online.)  Local DOS on the $A$ sublattice for $U=3$ at different average times with $E_0=1$. The same average times are chosen as shown in Fig.~\ref{fig: ldos_u=1.5}. In this case, the evolution of the DOS is much more mild, but we definitely see a reduction in the magnitude of the overall gap for long times and a splitting of the lower band, while the upper band does not appear to change too significantly.} \label{fig: ldos_u=3}
\end{figure}

The case $U=3$ and $E_0=1$ is plotted in Fig.~\ref{fig: ldos_u=3}.  It shows that the lower band can split and be shifted. Already at $t_{ave}=0$, the nonequilibrium local DOS shows a split lower band that eventually approaches a shift of the (split) band edge by $\pm E_0$ from its equilibrium location at $-U/2$. The upper band density of states doesn't shift at all and it shows a large enhancement at the lower band edge of the upper band (although, we don't believe the singularity remains, but cannot go far enough out in time to verify this). In this case, there aren't as many field related gaps observed, perhaps because the DOS is too small at larger frequencies for those structures to be seen. More intriguing is the fact that the structure tends to be more broadened at larger $U$ even though there still is no interaction in the system.

\begin{figure} 
\centerline{\includegraphics[width=100mm,clip=on]{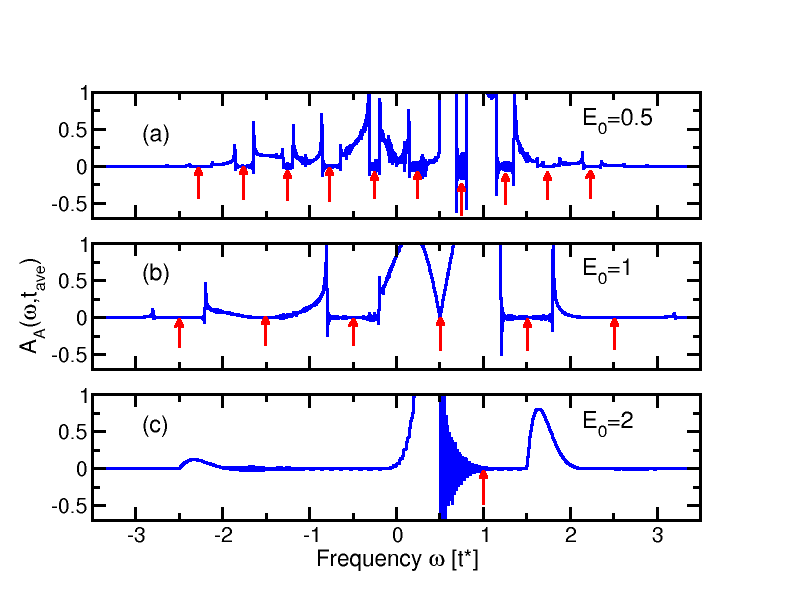}}
\caption{(Color online.) Evolution of the ``steady state''($t_{ave}=300$) local DOS for the $A$ sublattice and $U=1$  for different DC field amplitudes. Red arrows show the positions of the electric field induced gaps. } \label{fig: ldos_field}
\end{figure}

\begin{figure}
\centerline{\includegraphics[width=100mm,clip=on]{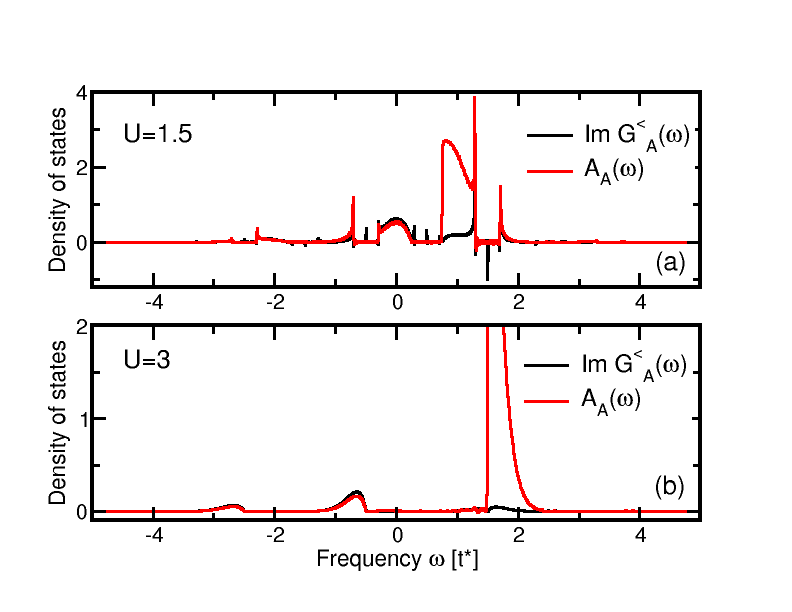}}
\caption{(Color online.) Imaginary part of the lesser Green's function (black) at (a) $ U=1.5$ and (b) $U=3$ compared to the corresponding local DOS (red)  on the $A$ sublattice with $E_0=1$. The curves where calculated at $t_{ave}=500$. The ratio of the black curve to the red curve gives the local distribution function, which would be Fermi-Dirac-like for a thermalized system. The distribution function that one would extract from the $U=1.5$ case is clearly nonthermal (focus on the region around $\omega=1$), since it would not be monotonic. It is more difficult to judge whether the case with $U=3$ could be described by some effective thermal distribution, but the last peak (near $\omega=2$) does not look like it has the right shape to be described by a Fermi-Dirac-like distribution. } \label{fig: dos_lesser}
\end{figure}

We next show the evolution of the local DOS on the $A$ sublattice for $U=1$ and various field amplitudes in Fig.~\ref{fig: ldos_field}. The arrows indicate the positions of field-induced gaps in the spectrum. We find that for all cases of the field amplitudes, that the DOS shows gaps at half-odd integer multiples of $E_0$, just like what we already observed in Fig.~\ref{fig: ldos_u=1.5} when we had $U=1.5$.  We also note that in some cases, there is no miniband centered around a specific gap that we expect to see. This occurs for large frequencies for all cases. In addition, the case with $E_0=1$ and $\omega=0.5$ [panel (b)] shows a pseudogap-like behavior, with the ``gap'' occurring at just one point. The field magnitude controls the size of the gap as well, as we see the gap size grow as $E_0$ increases. In the $E_0=2$ case, only the gap at $\omega=1$ is clear, while the other gap edges do not show significant features. One can still see a Wannier-Stark-like ladder in these systems, but the peaks no longer occur just at the Bloch frequencies, but are shifted, and also broadened, as we expect them to be. Clearly the behavior of the DOS is complex, and depends in a nontrivial way on the site energy $U$ and the field amplitude $E_0$. There doesn't appear to be any simple way to relate them to Wannier-Stark physics in a coherent fashion.

The imaginary part of the Fourier transform of the local lesser Green's function at $t_{ave}=500$ is plotted for two different $U$ values and compared to the local DOS in Fig.~\ref{fig: dos_lesser}. The ratio of the lesser curve to the DOS curve gives the local distribution function.  If the system was driven to a thermal distribution at long times, then the distribution function would be monotonic and given by the Fermi-Dirac distribution function with some effective temperature. This clearly is not the case for $U=1.5$, since one can see for the miniband centered around $\omega=1$ that the distribution function must be increasing for a finite range of frequency. It is more difficult to see whether the larger $U$ case could be described via the fluctuation-dissipation theorem, but the miniband occupancy near $\omega=2$ does not have the same shape as the local DOS, and hence it looks like this would require a nonmonotonic distribution function as well.

\section{Conclusions}

In this work, we have given details about how one can exactly solve for the Bloch-oscillation problem in a lattice with a basis, as given by the two-sublattice CDW order.  This is the simplest generalization of the Bloch oscillation problem to multiple bands and surprisingly, it has much richer behavior than one might have naively expected.  The problem can be solved exactly by working in a $2\times 2$ basis for the evolution operators for each momentum pair ${\bf k}$ and ${\bf k}+{\bf Q}$. We developed this formalism and showed how to calculate relevant observables.  We also verified a number of nontrivial exact relations that the solution must satisfy including moment sum rules of the local DOS and an equality of two different forms for the average energy as a function of time. This formalism works and is exact in any spatial dimension.  For concreteness, and to compare to work done for interacting systems, we solved the problem in infinite dimensions on the hypercubic lattice.

We used this formalism to study the Bloch-oscillation problem given by the question of how does this system evolve after a large amplitude field is turned on at $t=0$.  We found that the current undergoes a transient response before settling into a steady state-like behavior for longer times, but the time trace of the current versus time is not governed solely by the Bloch frequency but shows more complex structure and behavior.

We also examined the local DOS to see how the Wannier-Stark-ladder physics was modified by the gap in the spectrum.  We found complex behavior here as well, which is not determined solely by any simple rule for how the system will evolve, although we did find a propensity for minibands to form with gaps at half-odd integer multiples of the field amplitude.

Finally, we examined whether the fluctuation-distribution theorem holds at long times, and we found that typically, as one might expect, the systems do not follow a thermal distribution, but show a markedly athermal distribution of states.

The work we presented here will be challenging to observe in CDW systems in condensed matter, but they could be seen more easily in cold atom systems on double-well optical lattices, where this model is a natural model to describe the behavior of those systems. While the local DOS cannot be measured, analogs to photoemission experiments are possible, as is the possibility of measuring the number of particles in each band as a function of time.  It might even be possible to construct the current versus time by processing time-of-flight images to determine the momentum distribution functions and summing them over momentum (weighted by the velocity) to get the current.

\acknowledgments
The development of the parallel computer algorithms was supported by the National Science Foundation under Grant No. OCI-0904597. The data analysis was supported by the Department of Energy, Office of Basic Energy Research under Grants No. DE-FG02-08ER46542 (Georgetown), DE-AC02-76SF00515 (Stanford/SLAC), and DE-SC0007091 (for the collaboration). High performance computer resources utilized the National Energy Research Scientific Computing Center supported by the Department of Energy, Office of Science, under Contract No. DE- AC02-05CH11231. J.K.F. was also supported by the McDevitt bequest at Georgetown.

\appendix

\section{Calculation of first three moments of the retarded Green's function}

The $\emph{n}th$ order moments are defined as follows:
\begin{equation}
\mu_n^{ii}(t_{ave})=\int_{-\infty}^{+\infty}d \omega \omega^n\emph A_{ii}(\omega,t_{ave}).
\end{equation}
This expression is equivalent to
\begin{equation}
\mu_n^{ii}(t_{ave})= -{\rm Im}[i^n\frac{\partial^n}{\partial t_{rel}^n}G^{R}_{ii}(t_{rel},t_{ave})]_{t_{rel}=0^+}.
\label{eq: moment_deriv}
\end{equation}

We now verify the exact results for the first three moments for our retarded Green's function. In infinite dimensions, the moments for the local retarded Green's function in our inhomogeneous system satisfy the following \cite{sumrules}: (1) the zeroth order moment is 1; (2) the first order moment is $\pm U/2$; (3) the second order moment is $1/2+U^2/4$ ; and (4) the third order moment is $\pm U/4 \pm U^3/8$. (The A sublattice has a plus sign in the first and third moments). The zeroth moment follows directly by just setting $t\rightarrow t'^+$ in the retarded Green's function.
Since the retarded Green's functions are just linear combinations of the elements of the time-evolution operators, we can directly verify the moments by calculating the first three derivatives of the time evolution matrix (we set $\mu=U/2$ to make the formulas less cumbersome).

The first derivative is easy to find:
\begin{eqnarray}
\frac{\partial{U(k,t,t')}}{\partial t_{rel}}&=&-\frac{i}{2}\left(\begin{array}{cc}{\varepsilon_k(t)}\ \ \ \  {\frac{U}{2}}\\{\frac{U}{2}}\ \ \ \ {-\varepsilon_k(t)}\end{array}\right)U(k,t,t')\nonumber\\
&-&\frac{i}{2}U(k,t,t')\left(\begin{array}{cc}{\varepsilon_k(t')}\ \ \ \ {\frac{U}{2}}\\{\frac{U}{2}}\ \ \ \ {-\varepsilon_k(t')}\end{array}\right)
\label{eq: evolution_deriv}
\end{eqnarray}
(the factor of $1/2$ comes from the fact that $t=t_{ave}+t_{rel}/2$ and $t'=t_{ave}-t_{rel}/2$).
In the limit $t_{rel}\rightarrow 0^+$, $U(k,t,t)\rightarrow\textbf{I}$ and the first derivative becomes
\begin{equation}
\frac{\partial{U(k,t,t')}}{\partial t_{rel}}|_{t_{rel}\rightarrow 0^+}=-i\left(\begin{array}{cc}{\varepsilon_k(t)} & {\frac{U}{2}} \\ {\frac{U}{2}} &{-\varepsilon_k(t)}\end{array}\right).
\label{eq: moment_first_deriv2}
\end{equation}
We substitute this result into the definition of the local retarded Green's function in Eq.~(\ref{eq: g_ret_local}), which requires us to sum over momentum. Using the fact that $\sum_k 1=1/2$ for a summation over the reduced Brillouin zone, then yields
\begin{equation}
\frac{\partial{G^R_{ii}(t,t')}}{\partial t_{rel}}|_{t_{rel}\rightarrow 0^+}=-\sum_k \left (\varepsilon_k(t)-\varepsilon_k(t)\pm\frac{U}{2}\pm\frac{U}{2}\right  )
\end{equation}
which becomes $\mp U/2$ and shows that the first-order moment is $\pm U/2$ from Eq.~(\ref{eq: moment_deriv}) with $n=1$.

The second derivative of the retarded Green's function can be obtained by differentiating the first derivative
\begin{eqnarray}
\frac{\partial^2{U(k,t,t')}}{\partial t_{rel}^2}&=&-\frac{i}{2}\frac{\partial}{\partial t_{rel}}\left(\begin{array}{cc}{\varepsilon_k(t)} & {\frac{U}{2}} \\ {\frac{U}{2}} &{-\varepsilon_k(t)}\end{array}\right) U(k,t,t')\nonumber\\
&-&\frac{i}{2}\left(\begin{array}{cc}{\varepsilon_k(t)}\ \ \ \ {\frac{U}{2}}\\{\frac{U}{2}}\ \ \ \ {-\varepsilon_k(t)}\end{array}\right) \frac{\partial U(k,t,t')}{\partial t_{rel}}\nonumber\\
&-&\frac{i}{2}U(k,t,t')\frac{\partial}{\partial t_{rel}}\left(\begin{array}{cc}{\varepsilon_k(t')} & {\frac{U}{2}} \\ {\frac{U}{2}} &{-\varepsilon_k(t')}\end{array}\right)\nonumber\\
&-&\frac{i}{2}\frac{\partial U(k,t,t')}{\partial t_{rel}}\left(\begin{array}{cc}{\varepsilon_k(t')}\ \ \ \ {\frac{U}{2}}\\{\frac{U}{2}}\ \ \ \ {-\varepsilon_k(t')}\end{array}\right) .
\label{eq: evolution_second}
\end{eqnarray}
The derivative of the $U(k)$ matrix was evaluated in Eq.~(\ref{eq: evolution_deriv}).  The derivative of the Hamiltonian matrix becomes
\begin{equation}
\frac{\partial}{\partial t_{rel}}\left(\begin{array}{cc}{\varepsilon_k(t)} & {\frac{U}{2}} \\ {\frac{U}{2}} &{-\varepsilon_k(t)}\end{array}\right)=\frac{1}{2}{\bf E}(t)\cdot\boldsymbol{ \nabla}_k \varepsilon_k(t)\left(\begin{array}{cc}{1} & {0} \\ {0} &{-1}\end{array}\right)
\end{equation}
with a similar equation for the $t'$ matrix, but with $t\rightarrow t'$ and an overall sign change on the right hand side.
Here ${\bf E}(t)$ is the time-dependent electric field ${\bf E}(t)=-d{\bf A}(t)/dt$. 
So the first and third terms in Eq.~(\ref{eq: evolution_second}) cancel in the limit $t=t'^+$, and the rest of the terms give 
\begin{equation}
\frac{\partial^2{U(k,t,t')}}{\partial t_{rel}^2}|_{t_{rel}\rightarrow 0^+}=-\left (\varepsilon_k^2(t)+\frac{U^2}{4}\right )\textbf{I}.
\label{eq: moment_second_deriv1}
\end{equation}
So the second derivative of the Green's function becomes
\begin{eqnarray}
\frac{\partial^2{G^R_{ii}(t,t')}}{\partial t_{rel}^2}|_{t_{rel}\rightarrow 0^+}&=&{i}\sum_k \left ( \varepsilon_k^2(t)+\frac{U^2}{4}+\varepsilon_k^2(t)+\frac{U^2}{4}\pm 0\pm 0 \right )\nonumber\\
&=&i \left (\frac{1}{2}+\frac{U^2}{4}\right )
\label{eq: moment_second_deriv2}
\end{eqnarray}
in the limit as $t\rightarrow t'^+$, where we used the relation $\sum_k \varepsilon^2_k(t)=1/4$ for the infinite-dimensional density of states. So using Eq.~(\ref{eq: moment_deriv}) with $n=2$ shows that the second moment satisfies $\mu_2^{ii}=1/2+U^2/4$.

The third derivative of $U(k,t,t')$ has many more terms,
\begin{eqnarray}
\frac{\partial^3{U(k,t,t')}}{\partial t_{rel}^3}&=&-\frac{i}{4} {\bf E}(t)\cdot\boldsymbol{\nabla}_k \varepsilon_k(t)\left(\begin{array}{cc}{1} & {0} \\ {0} &{-1}\end{array}\right)\frac{\partial U(k,t,t')}{\partial t_{rel}}\nonumber\\ 
&-&\frac{i}{4} \frac{\partial}{\partial t_{rel}} {\bf E}(t)\cdot\boldsymbol{\nabla}_k \varepsilon_k(t)\left(\begin{array}{cc}{1} & {0} \\ {0} &{-1}\end{array}\right) U(k,t,t')\nonumber\\
&-&\frac{i}{4} {\bf E}(t)\cdot\boldsymbol{\nabla}_k \varepsilon_k(t)\left(\begin{array}{cc}{1} & {0} \\ {0} &{-1}\end{array}\right) \frac{\partial U(k,t,t')}{\partial t_{rel}}\nonumber\\
&-&\frac{i}{2}\left(\begin{array}{cc}{\varepsilon_k(t)}\ \ \ \ {\frac{U}{2}}\\{\frac{U}{2}}\ \ \ \ {-\varepsilon_k(t)}\end{array}\right) \frac{\partial^2 U(k,t,t')}{\partial t_{rel}^2}\nonumber\\
&+&\frac{i}{4}\frac{\partial U(k,t,t')}{\partial t_{rel}} {\bf E}(t')\cdot\boldsymbol{\nabla}_k \varepsilon_k(t') \left(\begin{array}{cc}{1} & {0} \\ {0} &{-1}\end{array}\right)\nonumber\\
&+&\frac{i}{4}U(k,t,t') \frac{\partial}{\partial t_{rel}} {\bf E}(t')\cdot\boldsymbol{\nabla}_k \varepsilon_k(t')\left(\begin{array}{cc}{1} & {0} \\ {0} &{-1}\end{array}\right)\nonumber\\
&-&\frac{i}{2}\frac{\partial^2 U(k,t,t')}{\partial t_{rel}^2}\left(\begin{array}{cc}{\varepsilon_k(t')}\ \ \ \ {\frac{U}{2}}\\{\frac{U}{2}}\ \ \ \ {-\varepsilon_k(t')}\end{array}\right)\nonumber\\
&+&\frac{i}{4}\frac{\partial U(k,t,t')}{\partial t_{rel}} {\bf E}(t')\cdot\boldsymbol{\nabla}_k \varepsilon_k(t')\left(\begin{array}{cc}{1} & {0} \\ {0} &{-1}\end{array}\right) .
\end{eqnarray}
In the limit $t_{rel}\rightarrow 0^+$, we substitute in the first and second derivatives at equal times from Eqs. (\ref{eq: moment_first_deriv2}) and (\ref{eq: moment_second_deriv1}) and set $t=t'$ in the rest. This yields
\begin{eqnarray}
\frac{\partial^3{U(k,t,t')}}{\partial t_{rel}^3}&=&\frac{i}{4}\frac{\partial}{\partial t}[{\bf E}(t)\cdot\boldsymbol{\nabla}_k\varepsilon_k(t)]\left (\begin{array}{cc}{1} & {0} \\ {0} &{-1}\end{array}\right)\nonumber\\
&-&\frac{1}{2}{\bf E}(t)\cdot\boldsymbol{\nabla}_k\varepsilon_k(t)\left (\begin{array}{cc}{1} & {0} \\ {0} &{-1}\end{array}\right)\left (\begin{array}{cc} {\varepsilon_k(t)} & {\frac{U}{2}}\\ {\frac{U}{2}} &{-\varepsilon_k(t)}\end{array}\right)\nonumber\\
&+&\frac{1}{2}{\bf E}(t)\cdot\boldsymbol{\nabla}_k\varepsilon_k(t)\left (\begin{array}{cc} {\varepsilon_k(t)} & {\frac{U}{2}}\\ {\frac{U}{2}} &{-\varepsilon_k(t)}\end{array}\right)\left (\begin{array}{cc}{1} & {0} \\ {0} &{-1}\end{array}\right)\nonumber\\
&+&i\left (\varepsilon_k^2(t)+\frac{U^2}{4}\right ) \left (\begin{array}{cc} {\varepsilon_k(t)} & {\frac{U}{2}}\\ {\frac{U}{2}} &{-\varepsilon_k(t)}\end{array}\right).
\label{eq: evolution_third_deriv}
\end{eqnarray}
Since we will be substituting this result into the formula for the retarded Green's function, where the $U_{11}(k)$ term always appears plus the $U_{22}(k)$ term and similarly for the $U_{12}(k)$ and $U_{21}(k)$ terms, the only nonvanishing contributions come from the off diagonal pieces of the last term in Eq.~(\ref{eq: evolution_third_deriv}).
Hence, the third derivative of the retarded Green's function is
\begin{eqnarray}
\frac{\partial^3{G_{ii}(t,t')}}{\partial t_{rel}^3}|_{t_{rel}\rightarrow 0^+}&=&\pm\sum_k\left [ U\varepsilon_k^2(t)+\frac{U^3}{4})\right ]\nonumber\\
&=&\pm \left (\frac{U}{4}+\frac{U^3}{8}\right ).
\end{eqnarray}
So the third moment satisfies $\mu_3^{ii}=\pm (U/4+U^3/8)$ with plus on the $A$ and minus on the $B$ sublattices.
Hence, we have now analytically verified the results for the first three moment sum rules. This is an important check that our formalism is correct. 

\section{Derivation of the formulas for the energy and the heating rate}

An additional check that we will use for the current formula is to verify energy conservation. The total energy $E_{tot}$ is found by simply taking the expectation value of the Hamiltonian (at half-filling with $\mu=U/2$)
\begin{eqnarray}
E_{tot}(t)&=&\sum_{k:\varepsilon_k\le 0}\Big [\varepsilon_k(t)\langle c_k^\dagger(t)c_k^{}(t)\rangle
+\frac{U}{2}\langle c_k^\dagger(t)c_{k+Q}^{}(t)\rangle\nonumber\\
&+&\frac{U}{2}\langle c_{k+Q}^\dagger(t)c_k(t)\rangle-\varepsilon_k(t)\langle
c_{k+Q}^\dagger(t)c_{k+Q}(t)\rangle\Big ].\nonumber\\
\label{eq: energy_total}
\end{eqnarray}
Using the Green's functions in the gauge, then yields
\begin{eqnarray}
E_{tot}(t)&=&-i\sum_{k:\varepsilon_k\le 0}\Big [\varepsilon_k(t)\{G_{11}^<(k,t,t)-G_{22}^<(k,t,t)\}\nonumber\\
&+&\frac{U}{2}\{G_{12}^<(k,t,t)+G_{21}^<(k,t,t)\}\Big ].\nonumber\\
\end{eqnarray}

The power pumped into the system is due to the accelleration of the electron along the electric field. A direct computation shows that
\begin{equation}
\frac{\partial E_{tot}(t)}{\partial t}=\langle\textbf{j}\rangle \cdot \textbf{E}(t).
\label{eq: power}
\end{equation}
We now show this identity analytically, using our results for the current and the total energy.
The equation of motion for the lesser Green's function follows from derivative of the respective evolution operator. We find the following four equations:
\begin{equation}
\partial_t G^<_{11}(k,t,t')=i\frac{U}{2}G_{12}^<(k,t,t)-i\frac{U}{2}G_{21}^<(k,t,t);
\end{equation}
\begin{eqnarray}
\partial_t G^<_{12}(k,t,t')&=&i\frac{U}{2}G_{11}^<(k,t,t)-i\frac{U}{2}G_{22}^<(k,t,t)\nonumber\\
&-&2i\varepsilon_k(t)G_{12}^<(k,t,t);
\end{eqnarray}
\begin{eqnarray}
\partial_t G^<_{21}(k,t,t')&=&-i\frac{U}{2}G_{11}^<(k,t,t)+i\frac{U}{2}G_{22}^<(k,t,t)\nonumber\\
&+&2i\varepsilon_k(t)G_{21}^<(k,t,t);
\end{eqnarray}
and
\begin{equation}
\partial_t G^<_{22}(k,t,t')=-i\frac{U}{2}G_{12}^<(k,t,t)+i\frac{U}{2}G_{21}^<(k,t,t),
\end{equation}
in the limit $t'\rightarrow t$.
Hence, the derivative of the total energy with respect to time becomes
\begin{equation}
\frac{\partial E_{tot}(t)}{\partial t}=-i\sum_{k:\varepsilon_k\le 0}[\partial_t \varepsilon_k(t)]\{G_{11}(k,t,t)-G_{22}(k,t,t)\}
\end{equation}
since the terms that don't involve the time derivative of the time-dependent bandstructure all cancel (which also follows from the Feynman-Hellman theorem).
Using the fact that $\partial_t\varepsilon_k(t)={\bf E}(t)\cdot\boldsymbol{\nabla}_k\varepsilon_k(t)$ and the final result for the expectation value of the current in Eq.~(\ref{eq: current}) yields Eq.~(\ref{eq: power}).
 This is a second stringent test that the formalism is correct.

Finally, we examine the filling in the transient upper and lower bands by determining the corresponding occupancy. Similar to the equilibrium case, the creation operators for the instantaneous upper and lower bands are written as,
\begin{equation}
c_{k+}^{\dagger}(t)=\alpha_k(t) c_{k}^\dagger(t)+\beta_k(t) c_{k+Q}^{\dagger}(t)
\end{equation}
\begin{equation}
c_{k-}^{\dagger}(t)=\beta_k(t) c_{k}^\dagger(t)-\alpha_k(t) c_{k+Q}^\dagger(t)
\end{equation}
with the corresponding annihilation operators being the hermitian conjugate.
Now the electron occupancy in the upper band is denoted by $n_+(t)$ and that of the lower band by $n_{-}(t)$. 
\begin{eqnarray}
n_+(t)&=&\sum_{k:\varepsilon_k\le 0}\langle c_{k+}^{\dagger}(t)c_{k+}^{}(t)\rangle\\ &=&\sum_{k: \varepsilon_k\le 0}\Big [ \alpha_k^2(t) \langle c_{k}^\dagger(t) c_{k}^{}(t)\rangle +\beta_k^2(t)\langle c_{k+Q}^\dagger(t) c_{k+Q}^{}(t)\rangle \nonumber\\
&+&\alpha_k(t) \beta_k(t)\langle c_{k}^\dagger(t) c_{k+Q}^{}(t)+c_{k+Q}^\dagger(t) c_{k}^{}(t)\rangle\Big ] \nonumber
\end{eqnarray}
and 
\begin{eqnarray}
n_-(t)&=&\sum_{k:\varepsilon_k\le 0}\langle c_{k-}^{\dagger}(t)c_{k-}^{}(t)\rangle\\ &=&\sum_{k: \varepsilon_k\le 0}\Big [ \beta_k^2(t) \langle c_{k}^\dagger(t) c_{k}^{}(t)\rangle +\alpha_k^2(t)\langle c_{k+Q}^\dagger(t) c_{k+Q}^{}(t)\rangle \nonumber\\
&-&\alpha_k(t) \beta_k(t)\langle c_{k}^\dagger(t) c_{k+Q}^{}(t)+c_{k+Q}^\dagger(t) c_{k}^{}(t)\rangle \Big ]. \nonumber
\end{eqnarray}
Here we have the time-dependent generalization of the instantaneous eigenvectors with
\begin{equation}
\alpha_k(t)=\frac{\frac{U}{2}}{\sqrt{2\left [\varepsilon_{k}^2(t)+\frac{U^2}{4}-\varepsilon_{k}(t)\sqrt{\varepsilon_{k}^2(t)+\frac{U^2}{4}}\right ]}}
\end{equation}
and
\begin{equation}
\beta_k(t)=\frac{-\varepsilon_{k}(t)+\sqrt{\varepsilon_{k}^2(t)+\frac{U^2}{4}}}{\sqrt{2 \left [\varepsilon_{k}^2(t)+\frac{U^2}{4}-\varepsilon_{k}\sqrt{\varepsilon_{k}^2(t)+\frac{U^2}{4}}\right ]}}.
\end{equation}
Note that the momentum-dependent occupancies are just given by the summand element for each $k$ and $k+Q$ pair. One immediately sees that the filling satisfies
\begin{equation}
n_+(t)+n_{-}(t)=\sum_{k:\varepsilon_k\le 0} \langle c_{k}^\dagger(t) c_{k}(t)+ c_{k+Q}^\dagger(t) c_{k+Q}(t)\rangle
\end{equation}
as it must.
Furthermore, the energy becomes
\begin{eqnarray}
E_{tot}(t)&=&\sum_{k: \varepsilon_k\le 0} \left  [n_{k+}(t)\sqrt{\varepsilon_k^2(t)+\frac{U^2}{4}}\right .\nonumber\\
&-&\left . n_{k-}(t)\sqrt{\varepsilon_k^2(t)+\frac{U^2}{4}}\right ],
\label{eq: energy_total2}
\end{eqnarray}
which follows from the instantaneous eigenenergies. We now directly show that Eq.~(\ref{eq: energy_total2}) is equivalent to Eq.~(\ref{eq: energy_total}).
Substituting in the values for $n_+(t)$ and $n_-(t)$ shows that the coefficient of the 
$\langle c_{k}^\dagger(t) c_{k}^{}(t)\rangle$ term is
\begin{eqnarray}
&~&[\alpha_k^2(t)-\beta_k^2(t)]\sqrt{\varepsilon_k^2(t)+\frac{U^2}{4}}\nonumber\\
&=&\frac{\sqrt{\varepsilon_k^2(t)+\frac{U^2}{4}}\left [ \frac{U^2}{4}-\frac{U^2}{4}-2\varepsilon_k^2(t)+2\varepsilon_k(t)\sqrt{\varepsilon_k^2(t)+\frac{U^2}{4}}\right ]}{2\left [ \varepsilon_k^2(t)+\frac{U^2}{4}-\varepsilon_k(t)\sqrt{\varepsilon_k^2(t)+\frac{U^2}{4}}\right ]}\nonumber\\
&=&\frac{2\varepsilon_k(t)\left [ \varepsilon^2_k(t)+\frac{U^2}{4}-\varepsilon_k(t)\sqrt{\varepsilon_k^2(t)+\frac{U^2}{4}}\right ]}{2\left [ \varepsilon_k^2(t)+\frac{U^2}{4}-\varepsilon_k(t)\sqrt{\varepsilon_k^2(t)+\frac{U^2}{4}}\right ]}\nonumber\\
&=&\varepsilon_k(t).
\end{eqnarray}
The coefficient of $\langle c_{k+Q}^\dagger(t) c_{k+Q}^{}(t)\rangle$ is just the negative of this and
the coefficients of $\langle c_{k}^\dagger(t) c_{k+Q}^{}(t)\rangle$ and $\langle c_{k+Q}^\dagger(t) c_{k}^{}(t)\rangle$ are equal and satisfy
\begin{eqnarray}
&~&\alpha_k(t)\beta_k(t)\sqrt{\varepsilon_k^2(t)+\frac{U^2}{4}}\nonumber\\
&=&\frac{\frac U2 \left [-\varepsilon_{k}(t)+\sqrt{\varepsilon_{k}^2(t)+\frac{U^2}{4}}\right ]\sqrt{\varepsilon_{k}^2(t)+\frac{U^2}{4}}}{\left [ \varepsilon_k^2(t)+\frac{U^2}{4}-\varepsilon_k(t)\sqrt{\varepsilon_k^2(t)+\frac{U^2}{4}}\right ]}\nonumber\\
&=&\frac{\frac U2 \left [\varepsilon_k^2(t)+\frac{U^2}{4}-\varepsilon_k(t)\sqrt{\varepsilon_k^2(t)+\frac{U^2}{4}}\right ]}{\left [ \varepsilon_k^2(t)+\frac{U^2}{4}-\varepsilon_k(t)\sqrt{\varepsilon_k^2(t)+\frac{U^2}{4}}\right ]}\nonumber\\
&=&\frac{U}{2}.
\end{eqnarray}
Which proves the result we needed to show and which provides the third stringent test of the formalism.

\end{document}